\documentstyle[12pt]{article}

\newcommand {\nn}    {\nonumber}
\newcommand {\vs}[1]  { \vspace*{#1 cm} }

\newcounter{eq}
\newcounter{sc}

\newcommand {\MPL}  {Mod.Phys.Lett.}
\newcommand {\NP}   {Nucl.Phys.}
\newcommand {\PL}   {Phys.Lett.}
\newcommand {\PR}   {Phys.Rev.}
\newcommand {\PRL}   {Phys.Rev.Lett.}
\newcommand {\CMP}  {Commun.Math.Phys.}

\def\overleftrightarrow#1{\vbox{\ialign{##\crcr
 $\leftrightarrow$\crcr\noalign{\kern-1pt\nointerlineskip}
 $\hfil\displaystyle{#1}\hfil$\crcr}}}

\newlength{\minitwocolumn}
\begin{document}


\begin{flushright}
EDO-EP-15\\
September, 1997\\
\end{flushright}
\vspace{30pt}

\pagestyle{empty}
\baselineskip15pt

\begin{center}
{\large\bf Space-time Uncertainty Principle from  \\
Breakdown of Topological Symmetry \vskip 1mm
}

\vspace{20mm}

Ichiro Oda
          \footnote{
          E-mail address:\ ioda@edogawa-u.ac.jp
                  }
\\
\vspace{10mm}
          Edogawa University,
          474 Komaki, Nagareyama City, Chiba 270-01, JAPAN \\

\end{center}


\vspace{15mm}
\begin{abstract}
Starting from topological quantum field theory,
we derive space-time uncertainty relation with respect to
the time interval and the spatial length proposed by
Yoneya through breakdown of topological symmetry 
in the large N matrix model.
This work suggests that the topological symmetry might
be an underlying higher symmetry behind the space-time
uncertainty principle of string theory.

\vspace{15mm}

\end{abstract}

\newpage
\pagestyle{plain}
\pagenumbering{arabic}


\rm
\section{Introduction}

In spite of recent remarkable progress in nonperturbative formulations
of M-theory \cite{M} and IIB superstring \cite{IKKT,FKKT}, we have
not yet reached a complete understanding of the fundamental principle
and the underlying symmetry behind string theory. Since a string
has an infinite number of states in the perturbative level in 
addition to various extended objects as solitonic excitations in 
the nonperturbative regime, it is expected that in string theory
the fundamental principle might have peculiar properties and the 
gauge symmetry would be quite huge compared to the usual ones in
particle theory.

In a quest of the fundamental principle of string theory,
Yoneya has advocated a space-time uncertainty principle
which is of the form \cite{Y1,Y2}
\begin{eqnarray}
\Delta T \Delta X 
\ge l_s^2,
\label{1.1}
\end{eqnarray}
where $l_s$ denotes the string minimum length which is related
to the Regge slope $\alpha'$ by $l_s = \sqrt{\alpha'}$.
Provided that the relation (1) holds literally, string
theory would lead a physical picture that space-time in itself 
is quantized at the short distance and the concept of space-time
as a continuum manifold cannot be extrapolated beyond the
fundamental string scale $l_s$. In more recent work \cite{Y2},
making use of the "conformal constraint" which stems from
the Schild action \cite{Schild} and essentially expresses
the space-time uncertainty priciple (1), Yoneya has
constructed a IIB matrix model from which the IKKT model
\cite{IKKT} can be interpreted as an effective theory
for D-branes \cite{Pol}.

In this article, we would like to consider the space-time 
uncertainty principle from a different perspective from
Yoneya's one. Namely, we attempt to understand at least 
some aspects of the underlying gauge symmetry of string 
theory through the relation (1). We will see
that topological quantum field theory \cite{Witten1} provides
us with a nice framework in understanding the space-time
uncertainty principle (1). It is worthwhile to point out
that it has been already stated that a topological symmetry 
might be of critical importance in both string theory and 
quantum gravity in  connection with the background independent
formulation of string theory and the unbroken phase of quantum 
gravity \cite{Witten2}. Our study at hand may shed a light 
on this idea to some extent.

The paper is organized as follows. In section 2 we review 
Yoneya's works \cite{Y2} relating to the present study.
In section 3, we derive the space-time uncertainty principle
from the topological field theory where the classical action 
is trivially zero. The final section is devoted to discussions.

\section{ The space-time uncertainty principle and \\
          the conformal constraint }

In this section, we review only a part of Yoneya's works relevant
to later study (See \cite{Y1,Y2} for more detail). Let us
start with the Schild action \cite{Schild} of a bosonic string.
Then the Schild action has a form
\begin{eqnarray}
S_{Schild} = -\frac{1}{2} \int d^2 \xi \left[\ -\frac{1}{2 \lambda^2}
\frac{1}{e} \left( \varepsilon^{ab} \partial_a X^\mu \partial_b
X^\nu \right)^2 + e \ \right],
\label{2.1}
\end{eqnarray}
where $X^\mu (\xi)$ $(\mu = 0, 1, \ldots , D-1)$ are space-time 
coordinates, $e(\xi)$ is a positive definite scalar density defined 
on the string world sheet parametrized by $\xi^1$ and $\xi^2$, 
and $\lambda = 4\pi \alpha'$. 

Taking the variation
with respect to the auxiliary field $e(\xi)$, one obtains
\begin{eqnarray}
e(\xi) = \frac{1}{\lambda} \sqrt{-\frac{1}{2} 
\left( \varepsilon^{ab} \partial_a X^\mu \partial_b
X^\nu \right)^2 },
\label{2.2}
\end{eqnarray}
which is also rewritten to be
\begin{eqnarray}
\lambda^2 = -\frac{1}{2}  
\left\{  X^\mu, X^\nu \right\}^2,
\label{2.3}
\end{eqnarray}
where one has introduced the diffeomorphism invariant Poisson
bracket defined as 
\begin{eqnarray}  
\left\{  X^\mu, X^\nu \right\} = 
\frac{1}{e(\xi)} \varepsilon^{ab} \partial_a X^\mu \partial_b
X^\nu.
\label{2.4}
\end{eqnarray}
Then eliminating the auxiliary field $e(\xi)$ through 
(3) from (2) and using the identity
\begin{eqnarray}  
- \det \partial_a X \cdot \partial_b X
= -\frac{1}{2} \left( \varepsilon^{ab} \partial_a X^\mu \partial_b
X^\nu \right)^2,
\label{2.5}
\end{eqnarray}
the Schild action (2) becomes at least classically 
equivalent to the famous Nambu-Goto action $S_{NG}$
\begin{eqnarray}
S_{Schild} &=& -\frac{1}{\lambda} \int d^2 \xi \sqrt{- \det 
\partial_a X \cdot \partial_b X},\nn\\
&=& S_{NG}.
\label{2.6}
\end{eqnarray}

Let us note that the "conformal" constraint (4) describes
half the classical Virasoro conditions \cite{Y2} and 
the well-known relation between the Poisson bracket and
the commutation relation in the large $N$ matrix model
\begin{eqnarray}
\left\{ A, B \right\} \longleftrightarrow 
\left[ A, B \right],
\label{2.7}
\end{eqnarray}
leads the "conformal" constraint (4) to
\begin{eqnarray}
\lambda^2 = -\frac{1}{2}  
\left[  X^\mu, X^\nu \right]^2.
\label{2.8}
\end{eqnarray}
Then it turns out that the commutation relation (9)
yields the space-time uncertainty principle (1) \cite{Y2}.
Here notice that it is not the whole Schild action but
the "conformal" constraint in the large $N$ matrix model that
produces the space-time uncertainty principle so that it is
a natural next step to seek the fundamental action yielding
the relation (9). Actually, Yoneya has derived such an action
which has a close connection with the IKKT model [5]. His 
construction of the action is in itself quite
interesting but seem to be a bit ambiguous. In particular,
a natural question arises whether or not we can derive the
relation (9) in terms of the more field-theoretic framework
where we usually start with a classical action with some 
local symmetry. In the following section, we shall challenge 
this problem of which we will see an interesting possibility
that the breakdown of a topological symmetry gives a
generation of the quantum action including the essential
content of the space-time uncertainty principle.

\section{ A topological model }

Let us start by considering a topological theory \cite{Witten1}
where the classical action is trivially zero but dependent on
the fields $X^\mu(\xi)$ and $e(\xi)$ as follows:
\begin{eqnarray}
S_{c} = S_{c}(X^\mu(\xi), e(\xi)) = 0.
\label{3.1}
\end{eqnarray}
The BRST transformations corresponding to the topological
symmetry are given by
\begin{eqnarray}
\delta_B X^\mu = \psi^\mu,  \ \delta_B \psi^\mu = 0, \nn\\
\delta_B e = e \ \eta,  \ \delta_B \eta = 0, \nn\\
\delta_B \bar{c} = b,  \ \delta_B b = 0,
\label{3.2}
\end{eqnarray}
where $\psi^\mu$ and $\eta$ are ghosts, and $\bar{c}$ and $b$
are respectively an antighost and an auxiliary field. Note that
these BRST transformations are obviously nilpotent. Also notice
that the BRST transformation $\delta_B e$ shows the character as a
scalar density of $e$.

The idea, then, is to fix partially the topological symmetry
corresponding to $\delta_B e$ by introducing an appropriate
covariant gauge condition. A conventional covariant and 
nonsingular gauge condition would be $e = 1$ but this gauge
choice is not suitable for the present purpose since it 
makes difficult to move to
the large $N$ matrix theory. Then almost unique choice up to
its polynomial forms is nothing but the "conformal" condition
(4). Hence the quantum action defined as $S_q = \int d^2 
\xi \ e L_q$ becomes
\begin{eqnarray}
L_q &=&  \frac{1}{e} \delta_B \left[ \bar{c} \left\{ e \left(
\frac{1}{2} \left\{ X^\mu, X^\nu \right\}^2 + \lambda^2 \right)
\right\} \right],\nn\\
&=&  b \left( \frac{1}{2} \left\{ X^\mu, 
X^\nu \right\}^2 + \lambda^2 \right)
- \bar{c} \left( \eta \left( -\frac{1}{2} \left\{ X^\mu, 
X^\nu \right\}^2 + \lambda^2 \right) 
+ 2 \left\{ X^\mu, X^\nu \right\} \left\{ X^\mu, \psi^\nu 
\right\} \right),
\label{3.3}
\end{eqnarray}
where the BRST transformations (11) were used.

What is necessary to obtain a stronger form of the space-time
uncertainty relation (9) is to move to the large $N$ matrix
theory where in addition to (8) we have the following
correspondences
\begin{eqnarray}
\int d^2 \xi \ e \longleftrightarrow Trace,\nn\\
\int {\it D} e \longleftrightarrow \sum_{n=1}^\infty,
\label{3.4}
\end{eqnarray}
where the trace is taken over $SU(n)$ group. These
correspondences can be justified by expanding the
hermitian matices by $SU(n)$ generators in the large
$N$ limit as is reviewed by the reference \cite{F}.
Here it is worth commenting one important point. As
in the IKKT model \cite{IKKT} the matrix size $n$ is
now regarded as a dynamical variable so that the
partition function includes the summation over $n$.
Even if the direct proof is missing, the summation
over $n$ is expected to recover the path integration
over $e(\xi)$. In fact, the authors of the reference
\cite{FKKT} have recently shown that the model of
Fayyazuddin et al. \cite{F} where a positive definite
hermitian matrix $Y$ is introduced as a dynamical
variable instead of $n$, belongs to the same 
universality class as the IKKT model \cite{IKKT}
owing to irrelevant deformations of the loop equation
\cite{FKKT}. Thus we think that the correspondences
(8) and (13) are legitimate even in the context
at hand.

Now in the large $N$ limit, we have
\begin{eqnarray}
S_q = Tr \left( b \left( \frac{1}{2} \left[ X^\mu, 
X^\nu \right]^2 + \lambda^2 \right)
- \bar{c} \left\{ \eta \left( - \frac{1}{2} \left[ X^\mu, 
X^\nu \right]^2 + \lambda^2 \right) 
+ 2 \left[ X^\mu, X^\nu \right] \left[ X^\mu, \psi^\nu \right]
\right\} \right).
\label{3.5}
\end{eqnarray}
Next by redefining the auxiliary field $b$ by $b + \bar{c} \ \eta$,
$S_q$ can be cast into a simpler form
\begin{eqnarray}
S_q = Tr \left( b \left( \frac{1}{2} \left[ X^\mu, 
X^\nu \right]^2 + \lambda^2 \right)
- 2 \lambda^2 \bar{c} \ \eta - 2 \bar{c} \left[ X^\mu, X^\nu \right] 
\left[ X^\mu, \psi^\nu \right] \right).
\label{3.6}
\end{eqnarray}
Then the partition function is defined as
\begin{eqnarray}
Z &=& \int {\it D}X^\mu {\it D}\psi^\mu {\it D}e {\it D}\eta 
{\it D}\bar{c} {\it D}b \ e^{- S_q},\nn\\
&=& \sum_{n=1}^\infty \int {\it D}X^\mu {\it D}\psi^\mu {\it D}\eta 
{\it D}\bar{c} {\it D}b \ e^{- S_q}.
\label{3.7}
\end{eqnarray}
At this stage, it is straightforward to perform the path integration
over $\eta$ and $\bar{c}$, as a result of which one obtains
\begin{eqnarray}
Z = \sum_{n=1}^\infty \int {\it D}X^\mu {\it D}\psi^\mu {\it D}b \
e^{- Tr \ b \ \left( \frac{1}{2} \left[ X^\mu, X^\nu \right]^2 
+ \lambda^2 \right)}.
\label{3.8}
\end{eqnarray}
In (17) there remains the gauge symmetry
\begin{eqnarray}
\delta \psi^\mu = \omega^\mu,
\label{3.9}
\end{eqnarray}
which is of course the remaining topological symmetry. Now let us
factor out this gauge volume or equivalently fix this gauge symmetry
by the gauge condition $\psi^\mu = 0$, so that the partition
function is finally given by  
\begin{eqnarray}
Z = \sum_{n=1}^\infty \int {\it D}X^\mu {\it D}b \
e^{- Tr \ b \ \left( \frac{1}{2} \left[ X^\mu, X^\nu \right]^2 
+ \lambda^2 \right)}.
\label{3.10}
\end{eqnarray}

It is remarkable that the variation with respect to the auxiliary
variable $b$ in (19) gives a stronger form of the space-time uncertainty
relation (9) and the theory is "dynamical" in the sense that
the ghosts have completely decoupled from (19). In other words, we
have shown how to derive the space-time uncertainty principle from
a topological theory through the breakdown of a topological symmetry
in the large $N$ matrix model. Why has the topological theory yielded
the nontrivial "dynamical" theory? The reason is very much simple.
In moving from the continuous theory (12) to the matrix theory (14),
the dynamical degree of freedom associated with $e(\xi)$ was replaced
by the discrete sum over $n$, on the other hand, the corresponding
BRST partner $\eta$ remains the continuous variable. This distinct
treatment of the BRST doublet leads to the breakdown of the topological
symmetry giving rise to a "dynamical" matrix theory. In this respect,
it is worthwhile to point out that while the topological symmetry
is "spontaneously" broken, the other gauge symmetries never be
violated in the matrix model (Of course, correctly speaking, 
these gauge symmetries reduce to the global symmetries in the
matrix model but this is irrelevant to the present argument.)
Moreover, notice that the above-examined phenomenon is a peculiar
feature in the matrix model with the scalar density $e(\xi)$,
which means that an existence of the gravitational degree of
freedom is an essential ingredient.

\section{ Discussions }

In this short article, we have investigated a possibility of the
space-time uncertainty principle advocated by Yoneya \cite{Y1,Y2}
to be derived from the topological field theory \cite{Witten1}.
The study at hand suggests that the underlying symmetry behind
this principle in string theory might be a topological symmetry as
mentioned before in a different context \cite{Witten2}. This rather
unexpected appearance of the topological field theory seems to be
plausible from the following arguments. Suppose that we live in the
world where the topological symmetry is exactly valid. Then we have
no means of measuring the distance owing to lack of the metric tensor
field so that there is neither concept of distance nor the space-time
uncertainty principle. If the topological symmetry, in particular, 
that associated with the gravitational field, is spontaneously
broken by some dynamical mechanism, an existence of the dynamical
metric together with a string would give us both concept of
distance and the space-time uncertainty principle. 

So far we have not paid attention to the number of the space-time
dimensions so much except the implicit assumption $D \ge 2$. An
intriguing case is $D = 2$ even if this specification is not
always necessary within the present formulation. In this special
dimension, the Nambu-Goto action which is at least classically 
equivalent to the Schild action as shown in (7) becomes not only
the topological field theory but also almost a surface term 
as follows:
\begin{eqnarray}
\sqrt{- \det \partial_a X \cdot \partial_b X} 
&=& \sqrt{- \left( \det \partial_a X^\mu \right)^2}, \nn\\
&=& \pm \det \partial_a X^\mu, \nn\\
&=& \mp \frac{1}{2} \varepsilon^{ab} \varepsilon_{\mu\nu}
\partial_a X^\mu \partial_b X^\nu,
\label{4.1}
\end{eqnarray}
where we have assumed a smooth parametrization of $X^\mu$ over
$\xi^a$ in order to take out the absolute value. Actually, this
topological model has been investigated to some extent in the 
past \cite{Fuji, Roberto, Oda}. In this case, it is interesting
that we can start with the nonvanishing surface term as a classical
action.

One of the most important problems in future is to understand 
the symmetry breaking mechanism of a topological symmetry 
proposed in this paper more clearly by physical picture. 
Another interesting problem is to introduce the spinors and 
construct a supersymmetric matrix model from the topological 
field theory. These problems will be reported in a separate 
publication.

\vs 1
\begin{flushleft}
{\bf Acknowledgement}
\end{flushleft}
The author thanks Y.Kitazawa and A.Sugamoto for valuable discussions. 
He is also indebted to M.Tonin for stimulating discussions and a
kind hospitality at Padova University where most of parts of this
study have been done. This work was supported in part by Grant-Aid 
for Scientific Research from Ministry of Education, Science and
Culture No.09740212.

\vs 1

\end{document}